\documentclass[prl,superscriptaddress,reprint]{revtex4-1}% Physical Review Letters aus preprint wird reprint
\usepackage{graphicx}% Include figure files
\usepackage{dcolumn}% Align table columns on decimal point
\usepackage{bm}% bold math
\usepackage[latin1]{inputenc}

\begin{document}

\title{Gain in Three-Dimensional Metamaterials utilizing Semiconductor Quantum Structures}

\author{Stephan Schwaiger}
\author{Matthias Klingbeil}
\author{Jochen Kerbst}
\author{Andreas Rottler}
\author{Ricardo Costa}
\author{Aune Koitm\"ae}
\author{Markus Br\"oll}
\author{Christian Heyn}
\author{Yuliya Stark}
\author{Detlef Heitmann}
\author{Stefan Mendach} \email{smendach@physnet.uni-hamburg.de}

\affiliation{%
Institut für Angewandte Physik und Zentrum für Mikrostrukturforschung, Universit\"at Hamburg, Jungiusstrasse 11, D-20355 Hamburg, Germany}%

\date{\today}

\begin{abstract}

We demonstrate gain in a three-dimensional metal/semiconductor metamaterial by the integration of optically active semiconductor quantum structures. The rolling-up of a metallic structure on top of strained semiconductor layers containing a quantum well allows us to achieve a three-dimensional superlattice consisting of alternating layers of lossy metallic and amplifying gain material. We show that the transmission through the superlattice can be enhanced by exciting the quantum well optically under both pulsed or continuous wave excitation. This points out that our structures can be used as a starting point for arbitrary three-dimensional metamaterials including gain.

\end{abstract}

\maketitle

Metamaterials are composite materials made of artificial building blocks whose size and lattice constant is small compared to the wavelength of the transmitted light. An advantage of metamaterials compared to conventional natural materials is that their properties can be tailored by varying the size and shape of the artificial building blocks. In particular using metallic structures, negative index of refraction in the near-infrared and visible regime had been realized using metallic split ring resonators~\cite{Linden2004} or fishnet structures~\cite{Dolling2006a,Dolling2007,Xiao2009}. To prove bulk properties and use metamaterials for devices one has to achieve a three-dimensional structure which is usually planar lithographically defined and fabricated sequentially until the desired thickness is reached \cite{Liu2007a,Liu2007,Valentine2008}. All these metamaterials are hampered by absorption which is caused by ohmic losses in the metallic compound or, in other words, by the finite imaginary part of the dielectric function of the metal. A proposal to compensate these losses is to integrate a gain medium into the metamaterial which amplifies the transmitted light \cite{Anantha2003a} like, e.g. dye \cite{Noginov2006} or semiconductor quantum structures~\cite{Govyadinov2007,Dong2010}. Recently it has been shown that a dye surrounding the metallic structures can compensate the losses in a metamaterial~\cite{Xiao2010a}. In contrast to dyes, semiconductor quantum structures exhibit no photo bleaching, a higher damage threshold and the possibility of electrical pumping. Investigations on two-dimensional planar systems show that metallic structures can be coupled to a quantum structure and partially compensate the ohmic losses~\cite{Vasa2008,Plum2009,Meinzer2010}.

In this letter, we present three-dimensional radial metamaterials consisting of alternating layers of metal structures and amplifying semiconductor quantum structures. For the fabrication we utilize the concept of self-rolling strained layers~\cite{Prinz2000,Schmidt2001,Schumacher2005} which we recently extended to strained metal/semiconductor metamaterials with a possible application as a hyperlens~\cite{Schwaiger2009,Smith2009}. Here we investigate a novel structure with active quantum wells integrated into the semiconductor layer. As a result we obtain a truly three-dimensional metamaterial containing optical amplifiers which enhance the light transmission by stimulated emission. We present transmission measurements which show a transmission enhancement of about 5~\% under optical continuous wave~(cw) excitation. Under pulsed excitation a relative enhancement in transmission of 11~\% was measured. We can well model our data by transfer matrix calculations, and we use finite-difference time-domain simulations~(FDTD) to illustrate for the example of the hyperlens how the measured gain improves the metamaterials' performance. 

\begin{figure}
\begin{center}
\includegraphics[scale=1]{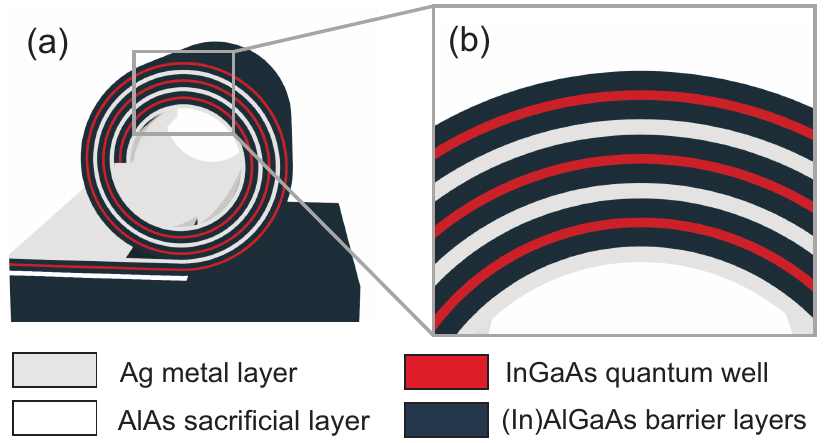}
\caption{(a)~A planar metal/semiconductor structure can be transformed into a three-dimensional metamaterial using the self-rolling concept of strained layers. (b)~The wall of the microtube is a metamaterial with alternating metallic~(Ag) and gain layers (InGaAs quantum well).}
\label{fig1}
\end{center}
\end{figure}

In Fig.~1(a) we present a sketch of our sample. Using molecular beam epitaxy (MBE) a GaAs buffer layer~(500 nm) is grown on a GaAs substrate, followed by an AlAs sacrificial layer (40 nm), the strained lower barrier layer consisting of Al$_{20}$In$_{13}$Ga$_{67}$As~(23 nm), the quantum well consisting of In$_{16}$Ga$_{84}$As~(7 nm) and finally the unstrained top barrier layer consisting of Al$_{23}$Ga$_{77}$As~(21 nm). Subsequently the structure is metalized with Ag~(13 nm) using thermal evaporation. By selective etching with HF the sacrificial layer is removed causing the strained layers to bend up while minimizing the strain energy. The wall of the microtube represents a superlattice of metallic and gain layers (Fig.~1(b)). All dimensions are more than one order of magnitude smaller than the wavelength of the transmitted light $\lambda>894\,$nm fulfilling the criteria for an effective metamaterial. 

\begin{figure}
\begin{center}
\includegraphics[scale=1]{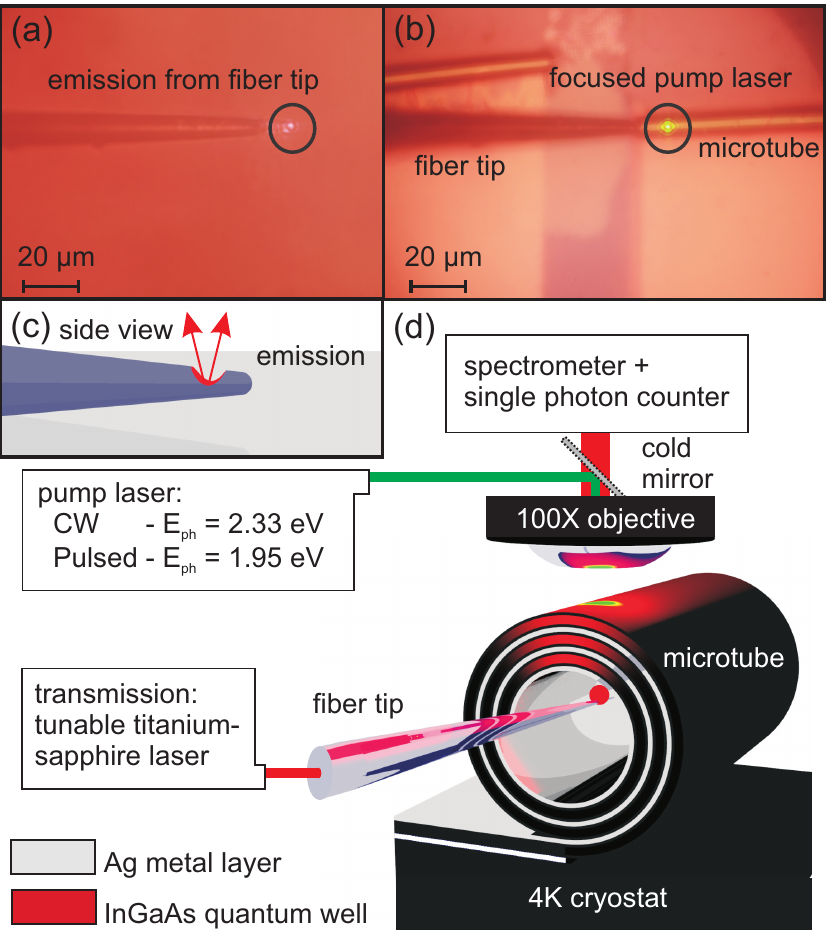}
\caption{(a)~Micrograph of a tapered optical fiber which is emitting light only from a spot on the side wall of the tip. (b)~The fiber tip can be inserted into a mircotube and illuminate it from the inside. To excite the quantum well in the tubes' wall, a pump laser can be focused onto the microtubes. (c)~Sketch of the side view of the light emitting fiber. (d)~To measure the transmission enhancement fiber and tube are located in a transmission setup. As a pump laser either a pulsed or a cw laser are used. Light from a tunable Ti:Sapphire laser is fed into the fiber and allows the transmission measurement.} 
\label{fig2}
\end{center}
\end{figure}

To measure the change in transmission through the radial metamaterial we used a tapered optical fiber, which was metalized and structured using focused ion beams. This fiber emits light from a hole in the side wall of the tip perpendicular to its axis and acts as a 'torch' (Fig.~2(a)) which can be placed inside a microtube (Fig.~2(b)). The microtube and the fiber tip are located inside a flow cryostat and can be moved with respect to each other using an XYZ piezo cube (attocube systems). Either a frequency doubled Nd:YAG cw laser ($E_{\mathrm{laser}}=2.33\,$eV) or a pulsed picosecond diode laser ($E_{\mathrm{laser}}=1.95\,$eV) can be focused with a spot size of $A_{\mathrm{spot}}\approx4\,\mu$m$^2$ onto the tube with a microscope objective (Fig.~2(d)). The transmitted light as well as the PL from the quantum well are collected with the same objective, energetically filtered with a spectrometer, detected with a single photon counting module, and acquired using a counter. For the measurements with the pulsed laser a start-stop-electronic is set between the counting module to only count events within a range of $\Delta t=10\,$ns around the laser pulse. The repetition rate of the laser was 1 MHz and the nominal pulse width $\tau_{\mathrm{FWHM}}=68\,$ps. In order to keep the pulse shape constant the power was adjusted by turning a lambda-half waveplate mounted between two crossed polarisation filters. Due to the complicated setup and our limited time resolution the peak power and pulse shape are not well characterized when incident onto the microtube. Nevertheless we also achieved valuable informations from the pulsed measurements. 

Since the transmitted light and the photoluminescence~(PL) exhibit the same energy we have to distinguish between transmission enhancement and PL as follows. Using an appropriate chopping system we acquire at a frequency of 150 Hz the dark level~(I$_{\mathrm{D}}$), the transmitted signal without pumping~(I$_{\mathrm{T}}$), the PL without the fiber emitting light~(I$_{\mathrm{PL}}$), and subsequently the transmitted signal along with the PL~(I$_{\mathrm{TPL}}$). All these values are taken within one period which allows very accurate measurements free from any long-term drift of the experimental setup. The relative transmission enhancement without PL: $\frac{\Delta T} {T}$=$\left(\frac{I_{\mathrm{TPL}}-I_{\mathrm{PL}}}{I_{\mathrm{T}} - I_{\mathrm{D}}} \right)-1$ is plotted in the following measurements vs. the photon energy of the transmitted light with a standard deviation of each transmission value of $\sigma{\left(\frac{\Delta T} {T}\right)} \approx 0.4\,$\% and $\sigma{\left(\frac{\Delta T} {T}\right)} \approx 0.8\,$\% for the cw and the pulsed measurements, respectively.

\begin{figure}
\begin{center}
\includegraphics[scale=1]{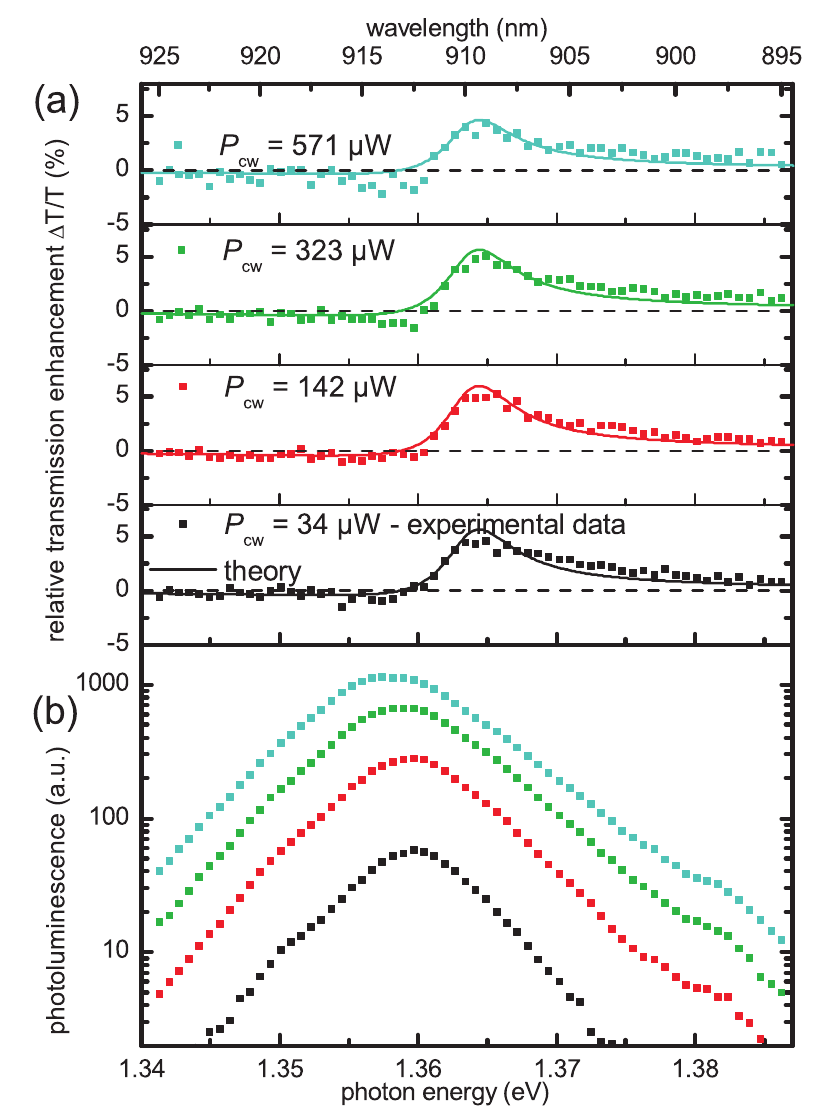}
\caption{(a)~Transmission enhancement spectra through the microtube under optical cw excitation at different pump powers (squares). At low pump powers (up to $P_{\mathrm{cw}}=323\,\mu$W) the enhanced transmission can be explained with a Lorentz oscillator model (solid lines). At high pump powers $P_{\mathrm{cw}}=571\,\mu$W deviations from the model occur. (b)~Corresponding PL data (squares). The PL exhibits a Lorentz shape. The peak position shifts to lower energy with increasing power due to the heating of the microtube.} 
\label{fig3}
\end{center}
\end{figure}

In Fig.~3(a) we present the transmission enhancement under cw excitation of a microtube which is rolled-up three times. At a pump power $P_{\mathrm{cw}}=34\,\mu$W (Fig.~3(a)) the transmission enhancement exhibits its maximum at $E_{\mathrm{T}}=1.365\,$eV. The maximum value is $\frac{\Delta T} {T}$$ = 4.6\,$\% with respect to the transmission if the quantum well is not pumped. At higher photon energies the transmission decreases towards its unpumped value. As will be discussed in more details later in the paper the transmission is slightly negative $\frac{\Delta T} {T}$$ \approx -0.7\,$\% at photon energies below $E_{\mathrm{T}}=1.359\,$eV. It increases again towards lower photon energies to its initial value.

The measured data can be confirmed by calculations using the transfer matrix method~\cite{Born1980} with the dielectric function $\epsilon$ for Ag and GaAs~\cite{Palik1985}. According to Ref.~\cite{Govyadinov2007} a Lorentz oscillator is added to the dielectric function of the semiconductor $\epsilon_{\mathrm{SC}}(\omega)$ to achieve the dielectric function of the gain layer $\epsilon_{\mathrm{Gainlayer}}(\omega)$:

\begin{equation}
		\epsilon_{\mathrm{Gainlayer}}(\omega) = \epsilon_{\mathrm{SC}}(\omega) + \left( \frac{A \omega_{\mathrm{L}}^2}  {\omega_{\mathrm{L}}^2 - \omega^2 - i \omega \gamma_{\mathrm{L}}} \right)
\end{equation}
where $A$ is an amplitude, $\omega_{\mathrm{L}}$ the central frequency of the quantum well and $\gamma_{\mathrm{L}}$ a damping frequency of the quantum well. We express the amplitude $A$ as a gain constant $\alpha$, which is the amplification per length, by the following equation:

\begin{equation}
		\alpha = - \frac{A \omega_{\mathrm{L}}^2}{Re(\sqrt{\epsilon_{\mathrm{SC}}})c_{0} \gamma_{\mathrm{L}}}	
\end{equation}
where $c_{0}$ is the speed of light. All spectra were normalised to the transmission spectra with a gain constant of $\alpha=0$.

The chosen Lorentz oscillator has its central energy at $E_{\mathrm{L}}=1.364\,$eV with a damping of $\gamma_{\mathrm{L}}=6\,$meV. The gain constant $ \alpha$ is fitted to each measurement. For the transmission enhancement at a power $P_{\mathrm{cw}}=34\,\mu$W we determine $\alpha=1,700\,$cm$^{-1}$. The corresponding PL is shown in Fig. 3b. It exhibits its maximum at $E_{\mathrm{PL}}=1.36\,$eV in the vicinity of the Lorentz oscillators' central energy indicating the quantum well as the origin of the transmission enhancement. The PL is red-shifted by $4$ meV with respect to the central energy of the Lorentzian, which indicates the well-known Stokes-Shift of the PL \cite{Bastard1984}. At higher excitation powers the PL increases in intensity and shifts to lower energies. At a power of $P_{\mathrm{cw}}=571\,\mu$W it is red-shifted by $2$ meV with respect to the low pumping-powers measurements. This effect can be explained by the increasing heating of the microtube and the related bandgap shift.

\begin{figure}
\begin{center}
\includegraphics[scale=1]{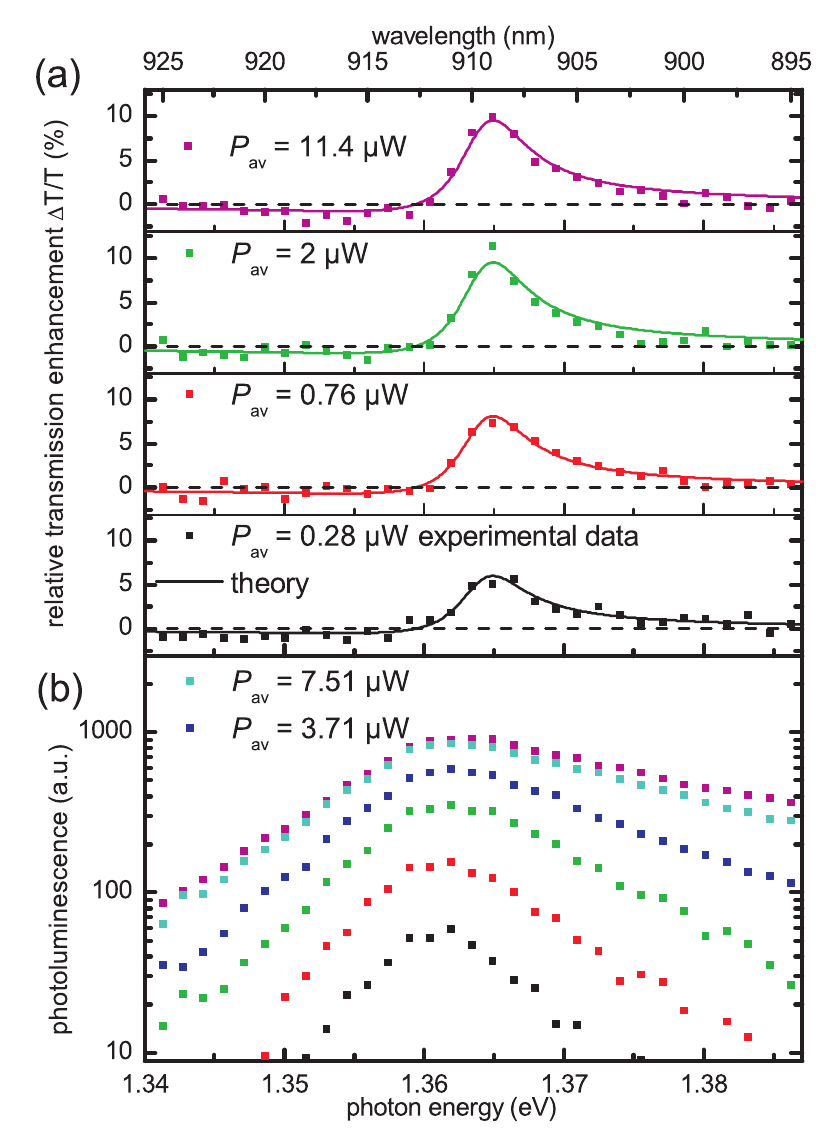}
\caption{(a)~Transmission enhancement spectra through the microtube under pulsed excitation at different pump powers (squares). Every measurement can be explained with the Lorentz oscillator model (solid lines) by assuming an increasing gain constant which saturates at $P_{\mathrm{av}}\approx 2\,\mu$W. (b)~The corresponding PL (squares) exhibits no red-shift but a high energy tail at pump powers above $P_{\mathrm{av}}=2\,\mu$W. This indicates that the lowest energy level in the quantum well is populated and stimulated emission occurs. For $P_{\mathrm{av}}=3.71\,\mu$W and $P_{\mathrm{av}}=7.51\,\mu$W only the PL data is shown.} 
\label{fig4}
\end{center}
\end{figure}

To exclude possible heating effects we also performed measurements with a pulsed laser which are shown in Fig.~4. At a power $P_{\mathrm{av}}=0.28\,\mu$W the shape of the transmission enhancement is comparable with the cw measurements. The correspoding calculations were performed with a gain constant of the Lorentz oscillator ($\alpha=1,800\,$cm$^{-1}$). By increasing the power $P_{\mathrm{av}}$ the maximum of the transmission enhancement increases and exhibits a value of $\frac{\Delta T} {T}$$ = 11.3\,$\% at $P_{\mathrm{av}}=2\,\mu$W. The corresponding gain constant is $\alpha=2,800\,$cm$^{-1}$. At excitation powers above $P_{\mathrm{av}}>2\,\mu$W the amplitude saturates at $\alpha=(2,780\pm130)\,$cm$^{-1}$. Even at the highest power, $P_{\mathrm{av}}=11.4\,\mu$W, no significant deviation from the model is observed. This clearly shows that the transmission enhancement can be attributed to the additional gain introduced into the dielectric function of the quantum well itself.

In Fig.~4(b) we present the corresponding PL data. All curves exhibit their maximum at the same energetical position as in the low power cw measurements (Fig.~3(b) $P_{\mathrm{cw}}=34\,\mu$W). The PL signal under pulsed excitation first increases with power but saturates at powers above $P_{\mathrm{av}}\approx7.5\,\mu$W. A clear difference compared to the cw measurements is the high energy tail of the PL at high pump powers. We explain this tail by band filling effects. A similar behavior in planar In(Ga)As quantum wells was observed in Refs.~\cite{Marcinkevicius1991,Tournie1993}. The bandfilling implies that the lowest energy levels are populated and the quantum well exhibits stimulated emission. Filling of the lowest levels also explains the saturation of gain observed at $P_{\mathrm{av}}>2\,\mu$W with pulsed excitation. The negative change in transmission in certain energy regimes, which is in the first moment surprising, can be attributed to a change of the dielectric function of the quantum well which changes the optical length of the metamaterial and therefore the Fabry-Pérot condition. The change in transmission in the high-power cw measurements ($P_{\mathrm{cw}}=571\,\mu$W) exhibits deviations from our model because our model does not include heating effects. Using a pulsed laser source we excluded heating effects while increasing the peak power with respect to the power $P_{\mathrm{cw}}$ in the cw measurements.

\begin{figure}
\begin{center}
\includegraphics[scale=1]{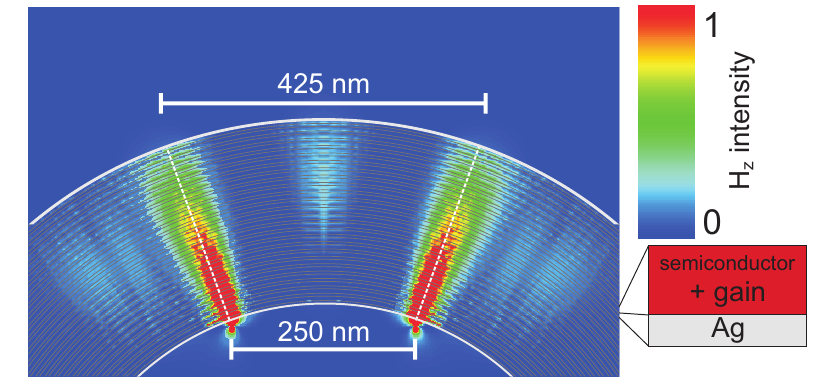}
\caption{FDTD simulation on the rolled-up metal/semiconductor superlattice containing 25 alternating layers of Ag ($d_{\mathrm{Ag}}=4\,$nm) and optically active semiconductor ($d_{\mathrm{SC}}=6\,$nm). The value $H_{z}$ is colour coded. Two dipoles are placed in a distance of 250 nm at the inner perimeter of the microtube. The electromagnetic waves emitted from the dipoles $E_{\mathrm{dipole}}=1.811\,$eV are transmitted radially through the superlattice allowing sub-wavelength imaging.} 
\label{fig5}
\end{center}
\end{figure}

To illustrate the influence of the gain on metamaterials we show in Fig. 5 the example of a hyperlens with gain. We performed FDTD simulations on a hyperlens similar to the one described in Ref. \cite{Schwaiger2009}. We placed two dipoles in a distance of 250 nm. The dipoles emit at the operation energy of the microtube at $E_{\mathrm{dipole}}=1.811\,$eV and they are placed at its inner perimeter. The microtube contains of 25 layers of Ag ($d_{\mathrm{Ag}}=4\,$nm) and optical active semiconductor ($d_{\mathrm{SC}}=6\,$nm and $\epsilon_{\mathrm{SC}}=14.1$). The gain constant of the semiconductor was chosen according to the measurements $\alpha=2,800\,$cm$^{-1}$. The electromagnetic waves are channeled through the wall of the microtube allowing their images to be resolved at the outer perimeter. By comparing this simulation with a simulation on a microtube with a passive semiconductor we found, that the functionality of the hyperlens remains while the transmission increases by $\frac{\Delta T} {T}$$ = 14\,$\%.

In conclusion we showed that using the concept of self-rolling we can fabricate a three-dimensional metamaterial. Our rolled-up metamaterial contains several alternating layers of metal and robust semiconductor gain material while maintaining the effective medium approximation. We demonstrate a transmission enhancement of $\frac{\Delta T} {T}$$ = 4.6\,$\% and $\alpha=1,700\,$cm$^{-1}$ under cw excitation. For pulsed excitation we found $\frac{\Delta T} {T}$$ = 11\,$\% and $\alpha=2,800\,$cm$^{-1}$. The latter values are sofar limited by our experimental time resolution. The actual gain might be higher if extracted from measurements with higher time resolution. We showed that for the example of a hyperlens the gain layers improve its operation. In the next step, the concept opens the road to utilise plasmonic resonances in three-dimensional metamaterials consisting of metallic nanostructures sandwiched between gain layers.

The authors thank W. Hansen and C. Strelow for fruitful discussions and A. Stemmann for MBE sample growth. We gratefully acknowledge support from the Universit\"at Hamburg and the Deutsche Forschungsgemeinschaft (DFG) through GrK 1286.

\end{document}